\documentclass{article}
\usepackage{amsmath}

\newtheorem{theorem}{Theorem}
\newtheorem{acknowledgement}[theorem]{Acknowledgement}

\input{tcilatex}
\begin{document}

\title{Pricing Equity Default Swaps under an approximation to the CGMY L\'{e}%
vy Model}
\author{S. Asmussen \\
Dept. of Mathematical Sciences\\
Aarhus University\\
Ny Munkegade\\
DK-8000 Aarhus C, Denmark \and D. Madan \\
Robert H. Smith School of Business\\
Van Munching Hall\\
University of Maryland\\
College Park, MD 20742 \and M. Pistorius \\
Dept. of Mathematics\\
King's College London\\
Strand \\
London WC2R 2LS\\
U.K.}
\maketitle

\begin{abstract}
The Wiener-Hopf factorization is obtained in closed form for a phase type
approximation to the CGMY\ L\'{e}vy process. This allows, for the
approximation, exact computation of first passage times to barrier levels
via Laplace transform inversion. Calibration of the CGMY model to market
option prices defines the risk neutral process for which we infer the first
passage times of stock prices to 30\% of \ the price level at contract
initiation. These distributions are then used in pricing 50\% recovery rate
equity default swap (EDS) contracts and the resulting prices are compared
with the prices of credit default swaps (CDS). An illustrative analysis is
presented for these contracts on Ford and GM.

\begin{acknowledgement}
This work was partly completed while all three authors were visiting the
Isaac Newton Institute as participants in the \textquotedblleft Developments
in Quantitative Finance\textquotedblright\ program. We thank Bruno Dupire
and other seminar participants at Bloomberg. We also thank conference
participants of the Credit Conference at Edinburgh and the MSRI conference
in London.
\end{acknowledgement}

\begin{acknowledgement}
In addition Martijn Pistorius acknowledges support from the Nuffield
Foundation Grant NAL/00761/G.
\end{acknowledgement}
\end{abstract}

\section{Introduction}

The equity default swap contract has recently been introduced in the
financial markets. The contract pays a recovery proportion of a notional
amount at the time of the occurence of a specific equity event, up to a
maturity, that is typically $5$ years. The equity event is defined as the
first time the stock price drops below $30\%$ of the price prevailing at
contract initiation. The distribution of this first passage time is then of
critical importance in the valuation of the contract. These distributions
are known for the geometric Brownian motion model and results have also
recently been obtained (Davydov and Linetsky (2001), Campi and Sbuelz
(2005)) for the constant elasticity of variance (CEV) model. The prices of
European options under these models, however, are not consistent with
observed market option prices and such observations call into question the
resulting swap prices.

Observed market option prices are more closely matched by a variety of pure
jump L\'{e}vy process models for the evolution of the logarithm of the stock
price. Here we shall focus attention on the four parameter $CGMY$ L\'{e}vy
model introduced by Carr, Geman, Madan and Yor (2002). First passage times
for such L\'{e}vy processes require knowledge of the distribution of the
supremum and infimum of the process over a fixed time interval. The product
of the Laplace transforms of these distributions is known by the famed
Wiener-Hopf factorization. Identification of the law of the infimum and
supremum then flounders on the inability to analytically perform the
factorization. Again, it is known that if the L\'{e}vy process has only
one-sided jumps, say downwards, then the law of the supremum is known and
one may then use the Wiener-Hopf identity to deduce the law of the infimum.
Such strategies have been effectively pursued in Rogers (2000), Novikov,
Melchers, Shinjikashvili and Kordzakhia (2003), Khanna and Madan (2004) and
Chiu and Yin (2005). Processes with one sided jumps are dominated by those
with two sided jumps in their ability to explain option prices.

For L\'{e}vy processes with two sided jumps the double Laplace transform in
time and the level of the infimum and supremum have been derived by Nguyen
and Yor (2002) for some L\'{e}vy processes using bivariate integral
representations. However, the numerical inversion of these transforms
involves four integrals and is computationally quite involved. Here we
consider an alternative strategy for obtaining the first passage
distribution of two sided jump L\'{e}vy processes. To this end, we use
phase-type distributions, defined as the time until absorption of a finite
state continuous-time Markov process (See Asmussen (1992)). It is a well
known fact that this class comes close to being complete for obtaining first
passage probabilities of the type we consider.

We employ results from Asmussen, Avram and Pistorius (2004) to obtain closed
forms for the Laplace transform of first passage distributions for L\'{e}vy
processes when the L\'{e}vy measure has a phase type distribution. In
addition to providing tractable computational schemes, phase-type
distributions have the advantage of forming a dense class so that by
increasing the size of the state space, one can in principle get arbitrarily
close to a target density and thereby get arbitrarily good approximations of
the first passage probabilities.

A particularly important subclass of phase-type distributions is formed by
the hyperexponential distributions, i.e. finite mixtures of exponentials.
The underlying Markov process then chooses some initial state, stays there
an exponential time with rate depending on the state and goes directly to
the absorbing state. This subclass is appealing for the $CGMY$ L\'{e}vy case
as the target density is completely monotonic, i.e. a possibly continuous
mixture of exponentials, so that by a discretization procedure it is a limit
of (possibly scaled) hyperexponential densities. In practice, the fitting
can be done either by maximum likelihood (Asmussen, Nerman and Olsson
(1996)) as then one wants to capture the shape of the target density, or by
minimizing a distance measure paying specific attention to specific features
of interest. In particular, in first passage problems a good fit in the tail
is crucial, and hyperexponential approximations to Pareto tails have been
obtained in this way (Feldman and Whitt (1998)). This is the path we follow
in this paper. We approximate the L\'{e}vy density of the $CGMY$ process by
a hyperexponential distribution and we then use the first passage time of
the approximating \ process to determine the Laplace transform of the first
passage time distribution for the approximating L\'{e}vy process. One
dimensional Laplace transform inversions using the methods of Abate and
Whitt (1995) then give us both the survival probability and the first
passage density for this approximation. One may wish to evaluate the quality
of this approximation to an exact result for the $CGMY$ process. The latter
is however not computationally available at this time. One may obtain
alternative approximations by either solving partial integro differential
equations or by Monte Carlo simulation. The simulation would require the
development of appropriate importance sampling methodologies. We therefore
leave these questions for future research. The same considerations apply
with respect to constructing a priori bounds on the error with respect to
the exact result.

From the first passage density and survival probability one may determine
the quote on an equity default swap contract. We estimate the $CGMY$
parameters from the prices of options on Ford and GM over a three year
period and extract daily estimates of quotes on equity default swap
contracts using the above hyperexponential approximation. We compare these
prices with market quotes on the $CDS$ rates and we observe that these two
separate sets of prices for a credit event, indeed, correlate well.

The outline of the rest of the paper is as follows. Section 2 presents the
details for the construction of the Laplace transform of the first passage
time for an approximation to the $CGMY$ L\'{e}vy process. Section 3
describes how this transform is used in quoting on equity default swap
contracts. Section 4 presents the results of daily calibration of the $CGMY$
model to market option prices. Section 5 compares the resulting equity
default swap prices with their credit default swap counterparts. Section 6
concludes.

\section{The CGMY\ First Passage Time}

We wish to determine the first passage time of the $CGMY$ L\'{e}vy process
to various levels using a phase type distribution to approximate the
process. For the Laplace transform of the first passage time of the
approximating process, i.e. a process with a phase type L\'{e}vy measure, we
follow Asmussen, Avram and Pistorius (2004). Given that the L\'{e}vy measure
is completely monotone we may employ a very special phase type distribution
that approximates the L\'{e}vy measure as a mixture of exponentials. The
phase type distribution we employ therefore has a simple form. Starting from
an initial state we have an entrance into a number of states where we stay
until absorption. There are no transitions between states. Suppose the L\'{e}%
vy measure has the form 
\begin{eqnarray}
k(x) &=&\sum_{i=1}^{n}a_{i}e^{-\alpha _{i}x}\mathbf{1}_{x>0}  \notag \\
&&+\sum_{j=1}^{m}b_{j}e^{-\beta _{j}|x|}\mathbf{1}_{x<0}.  \label{LM}
\end{eqnarray}

Our phase type distribution for the positive jumps has $n+1$ states with the
generator matrix%
\begin{equation}
T^{(+)}=\left( 
\begin{array}{cccc}
-\alpha _{1} &  &  & 0 \\ 
& -\alpha _{2} &  &  \\ 
&  & . & 0 \\ 
&  &  & -\alpha _{n}%
\end{array}%
\right)  \label{PTD}
\end{equation}%
with unnormalized initial state probabilities proportional to $\frac{a_{i}}{%
\alpha _{i}}$ for the states $i=1,\cdots ,n$ and the vector of absorption
rates into the final state is 
\begin{equation}
t=(\alpha _{1},\alpha _{2},\cdots ,\alpha _{n}).  \label{AR}
\end{equation}

The characteristic exponent for the approximation now takes the special form%
\begin{eqnarray}
\kappa (s) &=&\log E\left[ \exp \left( sX_{1}\right) \right]  \notag \\
&=&\mu s+\lambda _{+}\sum_{i=1}^{n}\pi _{i}^{+}\left( \frac{\alpha _{i}}{%
\alpha _{i}-s}-1\right) +\lambda _{\_}\sum_{j=1}^{m}\pi _{j}^{-}\left( \frac{%
\beta _{j}}{\beta _{j}+s}-1\right) +\frac{\sigma ^{2}}{2}s^{2}  \notag \\
\lambda ^{+} &=&\sum_{i=1}^{n}\frac{a_{i}}{\alpha _{i}};\text{ }\lambda
_{\_}=\sum_{j=1}^{m}\frac{b_{j}}{\beta _{j}}  \label{ce} \\
\pi _{i}^{+} &=&\frac{a_{i}}{\lambda _{+}\alpha _{i}};\text{ }\pi _{j}^{-}=%
\frac{b_{j}}{\lambda _{\_}\beta _{j}}.  \notag
\end{eqnarray}%
Where the value of $\sigma ^{2}$ will depend on the diffusion approximation
we incorporate for the small jumps. This is described in greater detail
later. The drift $\mu $ is chosen to make the overall drift on the stock
price be \ the interest rate less the dividend yield.

Let $T_{x}$ be the first passage time over a level $x>0,$ the Laplace
transform of $T_{x}$ at the transform argument $a$%
\begin{equation*}
E\left[ e^{-aT_{x}}\right]
\end{equation*}%
is obtained in terms of the roots, with positive real part, of the equation 
\begin{equation}
\kappa (s)=a.  \label{roots}
\end{equation}

We observe that one may write 
\begin{equation}
\kappa (s)=\frac{p(s)}{q(s)}  \label{polyroot}
\end{equation}%
as a ratio of two polynomials and so we seek the roots of the polynomial
equation 
\begin{equation}
p(s)=aq(s).  \label{polyroot2}
\end{equation}

The polynomial $q(s)$ is 
\begin{equation}
q(s)=\prod_{i=1}^{n}(\alpha _{i}-s)\prod_{j=1}^{m}\left( \beta _{j}+s\right)
\label{polyq}
\end{equation}%
and 
\begin{equation}
p(s)=q(s)\left( \frac{\sigma ^{2}}{2}s^{2}+\mu s-\left( \lambda _{+}+\lambda
_{-}\right) -\lambda _{+}\sum_{i=1}^{n}\pi _{i}^{+}\alpha _{i}\frac{1}{%
s-\alpha _{i}}+\lambda _{\_}\sum_{j=1}^{m}\pi _{j}^{-}\beta _{j}\frac{1}{%
\beta _{j}+s}\right) .  \label{polyp}
\end{equation}

We observe that $\kappa (0)=0$ and the poles of $\kappa (s)$ are exactly
equal to the eigenvalues of $-T^{(+)}$ and those of $T^{(-)}$ (and also the
roots of $q(s)=0$).We note that $p-aq$ is of degree $n+m+2$ if $\sigma >0$
and of degree $n+m+1$ if $\sigma =0,\mu >0.$ In the case $\mu >0,$ $\sigma
=0,$ $\kappa (s)=a$ has $n+1$ distinct positive roots and $m$ distinct
negative roots. If $\sigma >0$ $\kappa (s)=a$ has $n+1$ distinct positive
roots and $m+1$ distinct negative roots. Finally for $\mu =0,\sigma =0$
there are $n$ positive and $m$ negative distinct roots.

Let $k_{+}$ be the number of positive roots. Suppose the positive roots are $%
\rho _{i},$ $i=1,\cdots ,k_{+}$ then letting 
\begin{equation}
M_{t}=\sup_{s\leq t}\left( X_{s}\vee 0\right)  \label{sup}
\end{equation}%
and writing%
\begin{equation}
P\left( M_{e(a)}\in dx\right) =\int_{0^{+}}^{\infty }ae^{-at}P(M_{t}\in dx)
\label{supdistr}
\end{equation}%
($e(a)$ is an independent exponential random variable with mean $a^{-1}$),
we have that (for $s$ with $\mathcal{R}(s)\leq 0$)

\begin{eqnarray}
\phi _{a}^{+}(s) &=&\int_{0}^{\infty }ae^{-at}E\left[ e^{sM_{t}}\right] dt 
\notag \\
&=&P\left( M_{e(a)}=0\right) +\int_{0^{+}}^{\infty }e^{sx}P\left(
M_{e(a)}\in dx\right)  \notag \\
&=&\phi _{a}^{+}(-\infty )+\int_{0^{+}}^{\infty }e^{sx}P\left( M_{e(a)}\in
dx\right) .  \label{phiplus}
\end{eqnarray}%
This function may be expressed in terms of the positive roots $\rho _{i}$ by 
\begin{eqnarray}
\phi _{a}^{+}(s) &=&\frac{\det \left( -sI-T\right) }{\det \left( -T\right) }.%
\frac{\prod_{i=1}^{k_{+}}\left( -\rho _{i}\right) }{\prod_{i=1}^{k_{+}}%
\left( s-\rho _{i}\right) }  \notag \\
&=&\frac{\prod_{i=1}^{n}\left( 1-s\alpha _{i}^{-1}\right) }{%
\prod_{i=1}^{k_{+}}\left( 1-s\rho _{i}^{-1}\right) }  \label{phiplus2}
\end{eqnarray}%
where $\rho _{1},\cdots \rho _{k_{+}\text{ }}$are the positive roots of $%
\kappa (s)=a$ and where $T=diag(-\alpha _{i})$ is the restriction of $T^{+}$
to the nonabsorbing states. If $\sigma =0$ the number of positive roots $%
k_{+}$ is equal to the number of positive phases $n,$ $k_{+}=n$ and we note
that 
\begin{equation}
\phi _{a}^{+}(-\infty )=\prod_{i=1}^{k_{+}}\frac{\rho _{i}}{\alpha _{i}}
\label{phitail}
\end{equation}%
and $0<\phi _{a}^{+}(-\infty )<1.$

If $\sigma >0$ then the number of positive roots is $k_{+}=n+1$ one more
than the number of positive phases $n:$ $\rho _{1},\cdots \rho _{n+1}$ and
we have the formula 
\begin{equation}
\phi _{a}^{+}(s)=\frac{\prod_{i=1}^{n}\left( 1-s\alpha _{i}^{-1}\right) }{%
\prod_{i=1}^{n+1}\left( 1-s\rho _{i}^{-1}\right) }  \label{phiplus3}
\end{equation}%
furthermore, now $\phi _{a}^{+}(-\infty )=\lim_{s\rightarrow -\infty }\phi
_{a}^{+}(-s)$ is zero.

Hence in the case that there is a Brownian component present $(\sigma >0)$ 
\begin{equation}
\phi _{a}^{+}(-\infty )=0.  \label{phitail2}
\end{equation}

By performing a partial fraction decomposition we can perform the Laplace
inversion in $s$ analytically: Note that we can write 
\begin{equation}
\phi _{a}^{+}(s)-\phi _{a}^{+}(-\infty )=\sum_{i=1}^{k_{+}}A_{i}^{+}\frac{%
-\rho _{i}}{s-\rho _{i}}  \label{phiplus4}
\end{equation}%
where 
\begin{equation}
A_{i}^{+}=\frac{\prod_{j=1}^{n}\left( 1-\rho _{i}\alpha _{j}^{-1}\right) }{%
\prod_{j=1,j\neq i}^{k_{+}}\left( 1-\rho _{i}\rho _{j}^{-1}\right) }.
\label{Ai}
\end{equation}%
The Laplace transform can be inverted explicitly to find for $x>0$%
\begin{equation}
P\left( M_{e(a)}<x\right) =P\left( M_{e(a)}=0\right)
+\sum_{i=1}^{k_{+}}A_{i}^{+}(1-e^{-\rho _{i}x}).  \label{Mcdf}
\end{equation}%
It follows from the fact that 
\begin{equation}
P(M_{e(a)}<\infty )=1  \label{cdf1}
\end{equation}%
that 
\begin{equation}
1-\sum_{i=1}^{k_{+}}A_{i}^{+}=P\left( M_{e(a)}=0\right) =\phi
_{a}^{+}(-\infty ).  \label{phitail3}
\end{equation}%
In particular we get that

\begin{eqnarray}
P\left( M_{e(a)}>x\right) &=&1-P\left( M_{e(a)}=0\right)
-\sum_{i=1}^{k_{+}}A_{i}^{+}(1-e^{-\rho _{i}x})  \notag \\
&=&1-\sum_{i=1}^{k_{+}}A_{i}^{+}-P\left( M_{e(a)}=0\right)
+\sum_{i=1}^{k_{+}}A_{i}^{+}e^{-\rho _{i}x}  \notag \\
&=&\sum_{i=1}^{k_{+}}A_{i}^{+}e^{-\rho _{i}x}.  \label{Mccdf}
\end{eqnarray}

Now since 
\begin{eqnarray}
P\left( M_{e(a)}>x\right) &=&\int_{0}^{\infty }ae^{-at}P\left(
M_{t}>x\right) dt  \notag \\
&=&a\int_{0}^{\infty }e^{-at}P\left( T_{x}<t\right) dt  \label{fpt}
\end{eqnarray}%
we find the Laplace transform of $P\left( T_{x}<t\right) $%
\begin{equation}
\int_{0}^{\infty }e^{-at}P\left( T_{x}<t\right) dt=\frac{1}{a}%
\sum_{i=1}^{k_{+}}A_{i}^{+}e^{-\rho _{i}x}  \label{fptlt}
\end{equation}%
where $\rho _{i}=\rho _{i}(a)$ and $A_{i}^{+}=A_{i}^{+}(a)$ depend on $a.$

We employ the methods of Abate and Whitt (1995) to obtain first passage
probabilities and the first passage density. We next consider the details
for the diffusion approximation and the phase type approximation of the $%
CGMY $ process.

\subsection{CGMY details}

The $CGMY$ process was introduced in (Carr, Geman, Madan and Yor (2002), see
also Koponen (1995), Boyarchenko and Levendorskii (1999, 2000)) and is a
pure jump L\'{e}vy process with the L\'{e}vy density%
\begin{equation}
k(x)=C\frac{e^{-Mx}}{x^{1+Y}}\mathbf{1}_{x>0}+C\frac{e^{-G|x|}}{|x|^{1+Y}}%
\mathbf{1}_{x<0}.  \label{cgmylm}
\end{equation}%
For this completely monotone L\'{e}vy density of the $CGMY$ process we
recognize that 
\begin{equation}
\frac{1}{x^{1+Y}}=\int_{0}^{\infty }\frac{u^{Y}e^{-ux}}{\Gamma (1+Y)}du
\label{gamma}
\end{equation}%
and consider the approximation scheme 
\begin{equation}
\frac{1}{x^{1+Y}}\approx \sum_{i=1}^{N-1}\frac{%
u_{i}^{Y}e^{-u_{i}x}(u_{i+1}-u_{i})}{\Gamma (1+Y)}.  \label{fitfn}
\end{equation}

For a prespecified sequence of exponential decay coefficients $u_{i}$ that
correspond to reasonable levels of mean jump sizes under the single
exponential model. The specific values for $u_{i}$ were obtained using a
least squares algorithm that minimizes the sum of squared errors between the
left hand side of (\ref{fitfn}) and the right hand side of (\ref{fitfn})
evaluated at the $x$ points starting at $x=.25$ and increasing to $x=5$ in
steps of $.025.$ As starting values we employed $.5,2,5,10,20,40$ and $100$.
Our application works with the value of $Y$ fixed at $.5$ and hence the
approximation of equation (\ref{fitfn}) developed and subsequently used is
independent of all parameter variations. The resulting values for $u_{i}$
were $.1940,$ $.5982,$ $.8434,$ $1.1399,$ $1.5308,$ $2.1211,$ and $3.4055.$

We then approximate the $CGMY$ L\'{e}vy density on the two sides by 
\begin{eqnarray}
k_{+}(x) &=&\sum_{i=1}^{N-1}c_{i}e^{-(M+u_{i})x}  \label{cgmylmap} \\
k_{\_}(x) &=&\sum_{i=1}^{N-1}d_{i}e^{-(G+u_{i})x}.  \label{cgmylman} \\
c_{i} &=&d_{i}=\frac{Cu_{i}^{Y}(u_{i+1}-u_{i})}{\Gamma (1+Y)}
\end{eqnarray}

We then have in terms of our general discussion above%
\begin{equation}
\alpha _{i}=M+u_{i};\text{ }\beta _{i}=G+u_{i}  \label{coefs}
\end{equation}

Alternative but somewhat related procedures for fitting hyperexponentials to
general densities (rather than L\'{e}vy measures as here) have been
considered in Feldman and Whitt (1998), Andersen and Nielsen (1998) and
Asmussen, Jobmann and Schwefel (2002).

\subsection{First Passage to a low level}

For the first passage to a low level we may work with the first passage of $%
-X(t)$ to a high level. In this case we reverse the roles of $\alpha ,\beta $
and write 
\begin{eqnarray}
\alpha _{i} &=&G+u_{i}  \label{coefs2} \\
\beta _{i} &=&M+u_{i}.  \notag
\end{eqnarray}

\subsection{The small jump diffusion approximation}

Above we discussed a procedure to approximate a $CGMY$ process $X$ by a L%
\'{e}vy process $X_{1}$ with density (\ref{LM}) and to improve the
approximation we would like now to approximate the difference by a Brownian
motion.

A way to refine the above approximation of $X$ by $X_{1}$ is to approximate
the process of small jumps by a Brownian motion. This process of small jumps
denoted $Z_{1}^{\varepsilon }$ is obtained using the L\'{e}vy density of $X$
restricted to $(-\varepsilon ,\varepsilon ).$ The process $%
Z_{1}^{\varepsilon }$ is approximated by a Brownian motion with variance%
\begin{equation}
\sigma ^{2}=\sigma ^{2}(\varepsilon )=\int_{-\varepsilon }^{\varepsilon
}x^{2}\nu (dx).  \label{sigma}
\end{equation}

In Asmussen and Rosinski (2001) it is shown that, if $\nu $ has no atoms, $%
Z_{1}^{\varepsilon }$ weakly converges to a Brownian motion if and only if 
\begin{equation}
\lim_{\varepsilon \rightarrow 0}\frac{1}{\varepsilon }\int_{-\varepsilon
}^{\varepsilon }x^{2}\nu (dx)=\infty .  \label{int1}
\end{equation}

For a Variance Gamma process this diffusion approximation fails as

\begin{equation}
\lim_{\varepsilon \rightarrow 0}\frac{1}{\varepsilon }\int_{0}^{\varepsilon
}xe^{-ax}dx<\infty .  \label{int2}
\end{equation}

For a $CGMY$ process with $Y>0$ this diffusion approximation is valid, since%
\begin{equation}
\lim_{\varepsilon \rightarrow 0}\frac{1}{\varepsilon }\int_{0}^{\varepsilon
}x^{1-Y}e^{-ax}dx=\infty  \label{int3}
\end{equation}

(as $\nu (dx)/dx\geq const/|x|^{1+Y}$ in a neighbourhood of the origin, see
Example 2.3 in Asmussen and Rosinski (2001)).

In our $CGMY$ process approximation we truncated the $CGMY$ L\'{e}vy density
at $u_{1}=.5$ on the one hand but, on the other hand, the approximating
densities $k_{+}(x),k_{-}(x)$ are taken to start at $x=0.$ Therefore we
apply the diffusion approximation to 
\begin{equation}
\widetilde{k}(x)=\left( e^{-Mx}/x^{1+Y}-k_{+}(x)\right) \mathbf{1}%
_{0<x<u_{1}}+\left( e^{-G|x|}/|x|^{1+Y}-k_{\_}(x)\right) \mathbf{1}%
_{-u_{1}<x<0}.  \label{adjk}
\end{equation}

\section{Equity Default Swap Pricing}

We essentially follow the logic for the pricing of credit default swaps,
replacing the required survival probabilities and default time densities
with the first passage complementary distributions $\overline{F}%
_{n}(C,G,M,Y) $ and the first passage time density $f_{n}(C,G,M,Y)$ as
computed by Laplace transform inversion using the approximation methods
described above in section 2 for $\ n$ days or time $360t=n.$

Like the credit default swap, the equity default swap contract is viewed in
two parts, one describing the receipt side of the cash flows and the other
the payment side. There is a notional amount $M$ associated with the receipt
side and the actual level of cash flows received is $M$ times one minus the
recovery rate $R$ on the occurence of the equity event. Our calculations
employ a recovery rate of $50\%.$ These funds are received on the time $\tau 
$ of first passage of the equity to a level below the specified barrier. We
employ two barriers in our calculations, these are $50\%$ and $30\%$ of the
price at initiation.

Against this stream of receipts the equity default swap holder makes
periodic coupon payments, typically until the equity event or the maturity
whichever is comes first. At the first passage event date we subtract from
the receipts the accrued coupons from the last coupon date before $\tau $ to
this date. We denote time measured in days by $n$ and in years by $t.$ For
each day $n$ we define the function $\zeta (n)$ that gives the number of
days between the end of day $n$ and the last day on which a coupon was paid.
We also denote by $N$ the maturity of the contract in days while $T$ is the
maturity in years. By convention we take $(N,n)=360(T,t)$.

Let $\Delta (n)$ be the first passage event indicator function that takes
the value $1$ if the first passage of equity has occurred on or before day $%
n $ and is zero otherwise. Further, let $k_{T}$ denote the annual coupon
rate or the equity default swap rate quoted on the contract for maturity $T.$%
The cash flow receipts on day $n,$ $R_{n}$ are then written as 
\begin{equation}
R_{n}=\left( \Delta (n)-\Delta (n-1)\right) \left( M(1-R)-\frac{%
Mk_{T}(n-\zeta (n))}{360}\right) .  \label{Receipt}
\end{equation}

Let $NP$ denote the number of coupon payment dates and let $np_{j}$ be the
day number of the the $j^{th}$ payment date. The $j^{th}$ payment $P_{j}$
occurring on day $np_{j}$ is then given by 
\begin{equation}
P_{j}=\frac{Mk_{T}\left( np_{j}-np_{j-1}\right) }{360}\left( 1-\Delta
(np_{j})\right) ,  \label{Payment}
\end{equation}%
where the multiplication by the complementary first passage event indicator
function recognizes that no coupon payments are made after the first passage
event.

For the valuation of claims we employ a discount function $B_{n}$ that gives
the present value of a dollar promised on day $n.$ There are a variety of
approaches to constructing such discount functions.

The random present value of cash flows to the equity default swap contract, $%
V(T)$ is then given by 
\begin{equation}
V(T)=\sum_{n=1}^{N}R_{n}B_{n}-\sum_{j=1}^{NP}P_{j}B_{np_{j}}.
\label{SwapValue}
\end{equation}%
We are supposing here that conditional on the outcome of the first passage
event the expected values of future dollars are the same. This is equivalent
to supposing that the first passage events of single names in the economy do
not contain any information about macro movements in interest rates or that
interest rate evolutions are independent of the first passage process. We
leave for future research the modeling of joint evolutions of interest rates
and first passage times.

The equity default swap quotes in markets are set at levels consistent with
a zero price at the initiation of the swap contract. Hence we have that
under the risk neutral measure%
\begin{equation}
E^{Q}\left[ V(T)\right] =0.  \label{ZeroVal}
\end{equation}%
From our risk neutral distribution for the first passage time we employ 
\begin{equation}
E^{Q}\left[ (1-\Delta (n))\right] \underset{Def}{=}\overline{F}_{n}(C,G,M,Y).
\label{Survival}
\end{equation}%
The probability of first passage on a particular day $n$ may be approximated
by the density of default times the length of the day in years. 
\begin{eqnarray}
&&E^{Q}\left[ (\Delta (n)-\Delta (n-1))\right] \underset{Def}{=}\pi
_{n}(C,G,M,Y)  \notag \\
&\approx &f_{n}(C,G,M,Y)\frac{1}{365}.  \label{prob}
\end{eqnarray}

We may then compute the value of the equity default swap quote as 
\begin{eqnarray}
E^{Q}\left[ V(T)\right] &=&\sum_{n=1}^{N}\left[ M(1-R)-\frac{Mk_{T}(n-\zeta
(n))}{360}\right] B_{n}\pi _{n}(C,G,M,Y)-  \notag \\
&&\sum_{j=1}^{NP}\frac{Mk_{T}(np_{j}-np_{j-1})}{360}B_{np_{j}}\overline{F}%
_{np_{j}}(C,G,M,Y).  \label{SwapVal}
\end{eqnarray}%
Setting this value to zero and solving for the $k_{T}$ we obtain the $CGMY$
equity default swap pricing model 
\begin{equation}
k_{T}(C,G,M,Y)=\frac{(1-R)\sum_{n=1}^{N}B_{n}\pi _{n}(C,G,M,Y)}{%
\sum_{j=1}^{NP}\frac{(np_{j}-np_{j-1})}{360}B_{np_{j}}\overline{F}%
_{np_{j}}(C,G,M,Y)+\sum_{n=1}^{N}\frac{n-\zeta (n)}{360}B_{n}\pi
_{n}(C,G,M,Y)}.  \label{edsw}
\end{equation}%
Equation (\ref{edsw}) provides us with a four parameter equity default swap
pricing model and the parameters may be estimated from market prices of
options on the underlying name. In this way we obtain option implied quotes
for the equity default swap rates.

By way of a stylized example we set the interest rate yield curve at a flat $%
5\%$ level and used the values $C=.5,$ $G=2,$ $M=10,$ and $Y=.5$ with
recovery at $50\%$ and obtained the $EDS$ prices for maturities of $1,$ $3,$
and $5$ years at $161.97,$ $336.65$ and $439.54.$

\section{CGMY Calibration to Option Data}

We obtained daily data on $5$ year $CDS$ rates on Ford and GM over the
period 25 February 2002 to 25 February 2005. There were $696$ trading days
in this time interval. The credit default swap rates on these names saw a
sharp increase over this period before they finally came down again. The
mean CDS rates over this period were $308$ for Ford and $249$ for GM. The
standard deviations were $132$ and $78$ respectively while the minimum rates
were $155$ and $115$ respectively. The corresponding maximal rates were $720$
and $480.$

We employ the above algorithm to compute option implied equity default swap
rates with barriers for the equity event set at the typical value of $30\%$
of initiation price and a five year maturity. For this purpose we calibrate
the $CGMY$ model to European option prices for each day on each name. Equity
option prices on Ford and GM traded on the New York Stock Exchange are
typically available for American options. We mitigate the American feature
by first employing out of the money options. Second we infer implied
volatilities from American option prices using the Black Scholes model and
then construct the corresponding European prices by the Black Scholes
formula. The data is obtained from OptionMetrics and is available at $WRDS$
the Wharton Research Data Service.

We note that the case $Y=0$ is the variance gamma model, that successfully
calibrates any of the maturities. We are also aware that we may calibrate
equally well with any specific value of the parameter $Y.$ For our
approximations to have a diffusion component we require $Y>0$ and the
structure of the approximation is determined for any fixed value of $Y.$ We
therefore froze the value of the $Y$ parameter at $Y=.5$ and calibrated the
parameters $CGM$ for this frozen value of $Y.$ The calibrations were done
using maturities between $1$ and $2$ years, by minimizing the root mean
square error between market and model European option prices. \ We report in 
$Table$ $1$ summary statistics of the $CGM$ parameters for both the
companies.

\begin{equation*}
\begin{tabular}{lllllll}
\multicolumn{7}{l}{TABLE 1} \\ 
& \multicolumn{3}{l}{FORD} & \multicolumn{3}{l}{GM} \\ 
& C & G & M & C & G & M \\ 
Median & .6506 & 1.9458 & 11.0187 & .2171 & 1.0084 & 5.8031 \\ 
25\% & .3661 & 1.3066 & 9.9309 & .1664 & .6690 & 4.7486 \\ 
75\% & 1.0895 & 4.0969 & 11.3522 & .5582 & 2.7802 & 11.5872%
\end{tabular}%
\end{equation*}

\section{Equity Default Swap Rates and the CDS Rates}

We employed the calibrated $CGM$ model with $Y=.5$ to determine the equity
default swap rates for the typical barrier of $30\%$ for the payout on the
equity event. We present graphs of the implied EDS and CDS rates for both
companies over the 2002-2005 period evaluated every five days. The \ data on 
$CDS$ rates for the five year maturity are readily available from Bloomberg.

\FRAME{ftbpFU}{5.0194in}{4.0145in}{0pt}{\Qcb{Ford CDS in Blue, EDS in Red}}{%
}{edsvscdsfordbe.eps}{\special{language "Scientific Word";type
"GRAPHIC";maintain-aspect-ratio TRUE;display "USEDEF";valid_file "F";width
5.0194in;height 4.0145in;depth 0pt;original-width 6.7222in;original-height
5.188in;cropleft "0";croptop "1.0349";cropright "1";cropbottom "0";filename
'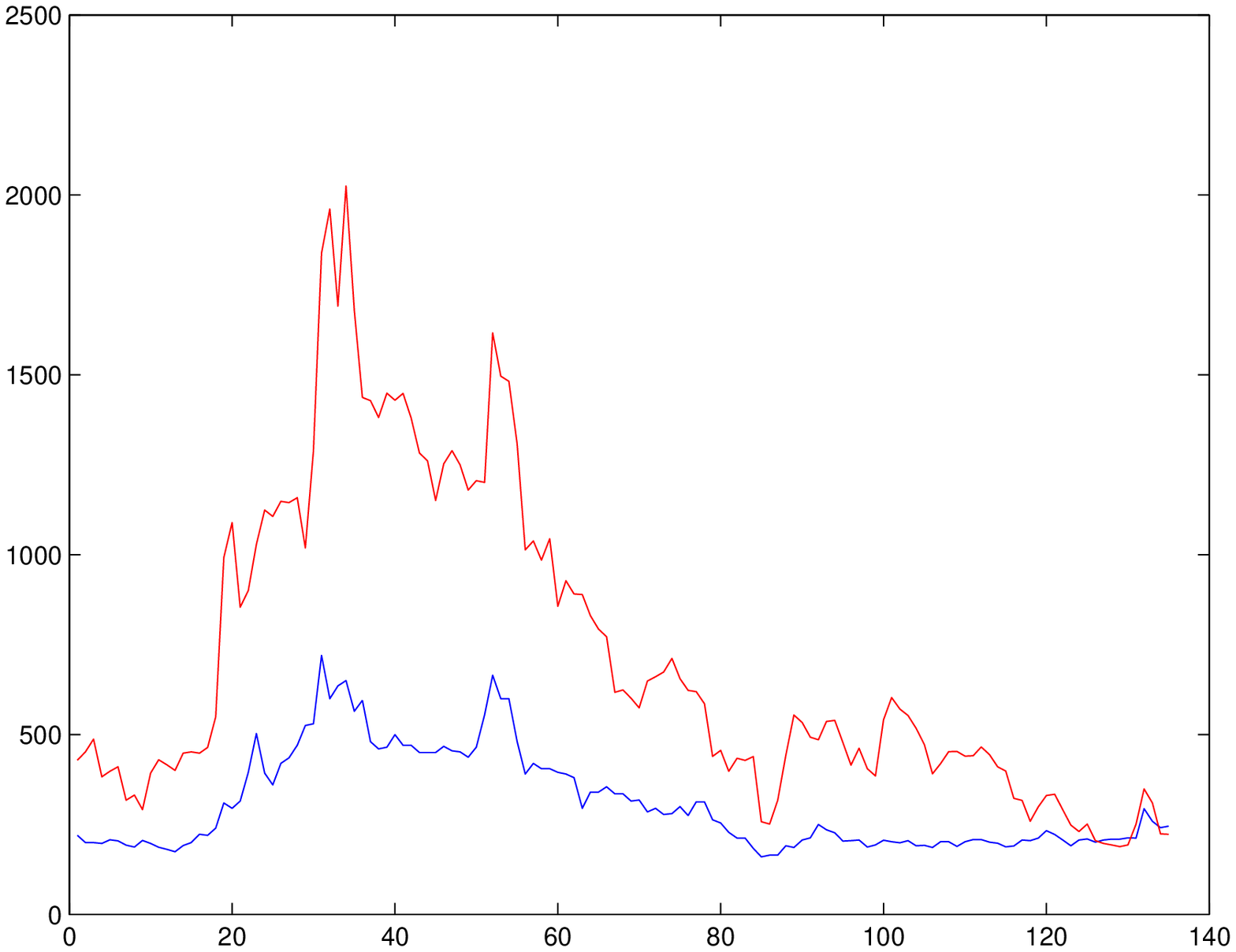';file-properties "XNPEU";}}

\bigskip

\FRAME{ftbpFU}{5.0194in}{4.0145in}{0pt}{\Qcb{GM CDS in Blue EDS in Red}}{}{%
edsvscdsgmbe.eps}{\special{language "Scientific Word";type
"GRAPHIC";maintain-aspect-ratio TRUE;display "USEDEF";valid_file "F";width
5.0194in;height 4.0145in;depth 0pt;original-width 6.7222in;original-height
5.188in;cropleft "0";croptop "1.0351";cropright "1";cropbottom "0";filename
'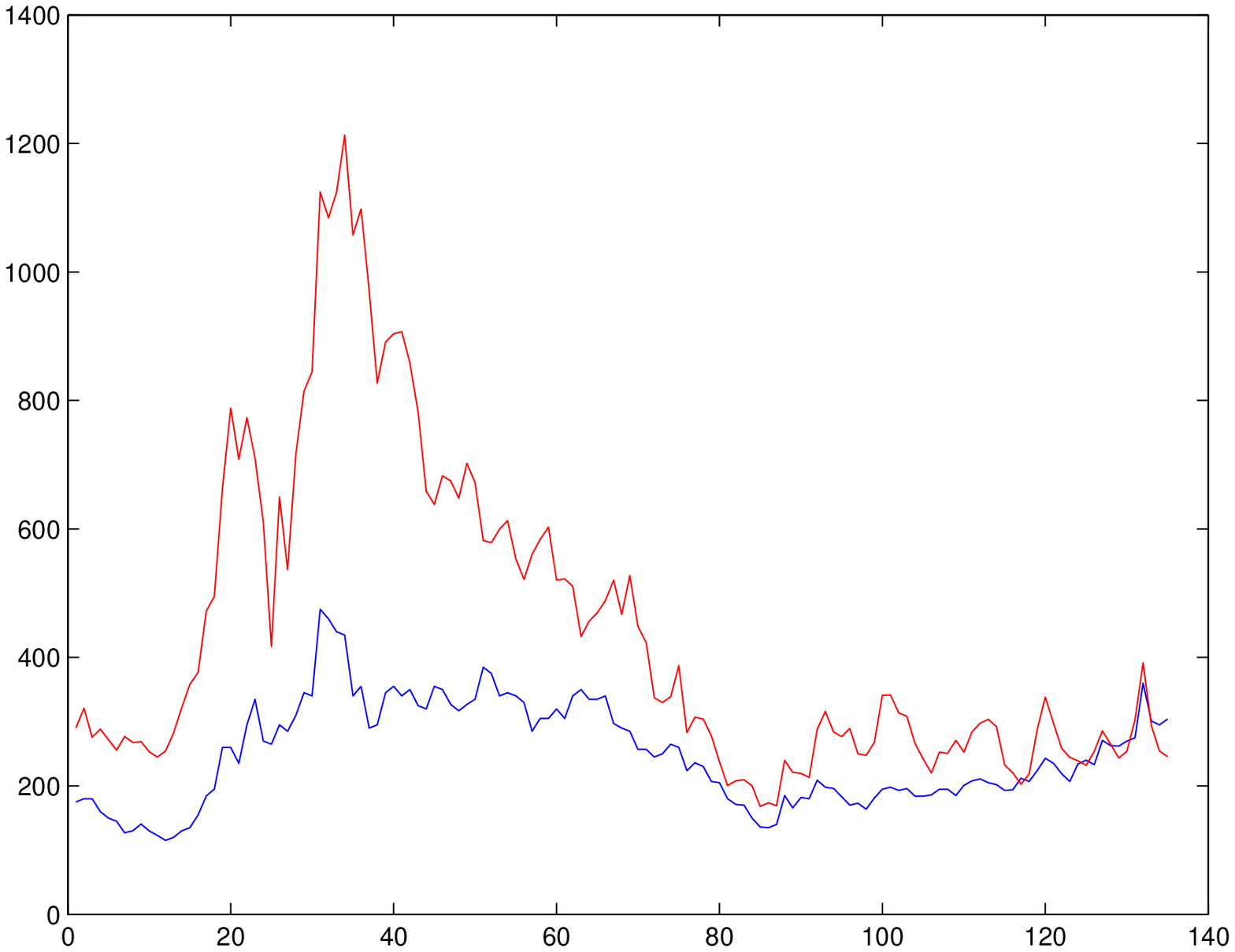';file-properties "XNPEU";}}

The results of regressing the CDS rates on the EDS rates are as follows. We
observe that there is a strong link between between the prices obtained from
these separate markets.%
\begin{equation*}
\begin{tabular}{lllllll}
& \multicolumn{3}{l}{$FORD$} & \multicolumn{3}{l}{$GM$} \\ 
& $Const$ & $Slope$ & $R^{2}$ & $Const$ & $Slope$ & $R^{2}$ \\ 
$value$ & $99.1820$ & $0.2913$ & $0.8950$ & $146.2816$ & $0.1442$ & $0.6292$
\\ 
$t-value$ & $13.62$ & $33.37$ &  & $18.10$ & $14.89$ & 
\end{tabular}%
\end{equation*}

\section{Conclusion}

Approximation of the $CGMY$ L\'{e}vy measure by hyperexponentials leads to
an exact Wiener-Hopf factorization of the approximating process and hence to
exact expressions for the first passage time of the approximating process to
a level. We then employ these results to obtain closed form formulas for the
prices of equity default swap contracts that payout on the equity event of a
loss of $70\%$ of the initial stock value with the receipts being a regular
coupon paid till the equity event or maturity, whichever is less.

The methods are illustrated on $CGMY$ processes calibrated to the vanilla
options market for $FORD$ and $GM$ over the period $25$ February $2002$ to $%
25$ February $2005.$ For the same period we also observe the daily values of
the credit default swap contracts and these are compared favorably with the
option imputed equity default swap prices computed by the $CGMY$
approximation method proposed here.

\bigskip \pagebreak

\end{document}